\documentclass[11pt]{article}
\usepackage{amsfonts,amsmath,amssymb}
\usepackage{enumerate}
\usepackage{hyperref}
\usepackage{bbm}
\usepackage{nicefrac}
\usepackage[all]{xy}
\usepackage{graphicx}
\usepackage{bm}
\usepackage{makecell}
\usepackage[table]{xcolor}

\usepackage{upgreek}

\usepackage{booktabs}

\newcommand{\be}{\begin{equation}}
\newcommand{\ee}{\end{equation}}

\newcommand{\bea}{\begin{eqnarray}}
\newcommand{\eea}{\end{eqnarray}}

\newcommand{\bes}{\begin{subequations}}
\newcommand{\ees}{\end{subequations}}

\newcommand{\cN}{{\cal N}}

\def\ft#1#2{{\textstyle{{\scriptstyle #1}\over {\scriptstyle #2}}}}

\def\sst#1{{\scriptscriptstyle #1}}

\def\0{{\sst{(0)}}}
\def\2{{\sst{(2)}}}
\def\3{{\sst{(3)}}}
\def\4{{\sst{(4)}}}
\def\5{{\sst{(5)}}}
\def\6{{\sst{(6)}}}
\def\7{{\sst{(7)}}}
\def\8{{\sst{(8)}}}

\newcommand{\vol}{\textrm{vol}}

\allowdisplaybreaks  

\usepackage{multirow}
\usepackage{rotating}

\newcommand{\ba}{\begin{align}}
\newcommand{\ea}{\end{align}}

\newcommand{\bse}{\begin{subequations}}
\newcommand{\ese}{\end{subequations}}

\usepackage{mathtools}
\usepackage{empheq}
\usepackage{cancel}

\renewcommand{\a}{{\alpha}}
\renewcommand{\b}{{\beta}}
\newcommand{\g}{{\gamma}}
\renewcommand{\d}{{\delta}}
\DeclareMathOperator{\e}{{\epsilon}}

\DeclareMathOperator{\ta}{\tilde{\alpha}}
\DeclareMathOperator{\tb}{\tilde{\beta}}

\DeclareMathOperator{\he}{\mathit{\hat{e}}}

\DeclareMathOperator{\hF}{\mathit{\hat{F}}}
\DeclarePairedDelimiter{\norm}{\Vert}{\Vert}

\DeclareMathOperator{\te}{\mathit{\tilde{e}}}

\renewcommand{\1}{\mathbbm{1}}

\usepackage{tensind}				
\tensordelimiter{?}

\DeclareMathOperator{\dd}{d\!}
\DeclareMathOperator{\grad}{\mathit{\nabla}}

\renewcommand{\L}{\mathcal{L}}

\begin{document}

\makeatletter
\renewcommand{\theequation}{\thesection.\arabic{equation}}
\@addtoreset{equation}{section}
\makeatother

\begin{titlepage}

\begin{flushright}
IFT-UAM/CSIC-19-107 \\
%
\end{flushright}

\vspace{20pt}

   \begin{center}
   \baselineskip=16pt

   \begin{Large}\textbf{
   Minimal $D=4$ ${\cal N}=2$ supergravity from $D=11$: \\[6pt]
   An M-theory free lunch}
   \end{Large}

\vspace{25pt}

{\large  Gabriel~Larios$^1$ \, and \,  Oscar Varela$^{1,2}$ }

\vspace{30pt}

	\begin{small}
          
  $^1$ {\it Departamento de F\'\i sica Te\'orica and Instituto de F\'\i sica Te\'orica UAM/CSIC , \\
   Universidad Aut\'onoma de Madrid, Cantoblanco, 28049 Madrid, Spain} 

		\vspace{15pt}
	
  $^2$  {\it Department of Physics, Utah State University, Logan, UT 84322, USA}

	\end{small}

\vskip 60pt

\end{center}

\begin{center}
\textbf{Abstract}
\end{center}

\begin{quote}

We present a new consistent truncation of $D=11$ supergravity to $D=4$ $\mathcal{N}=2$ minimal gauged supergravity, on the seven-dimensional internal Riemannian space corresponding to the most general class of $D=11$ solutions with an AdS$_4$ factor and $\cN=2$ supersymmetry. A truncation ansatz is proposed and its consistency checked at the level of the $D=11$ Bianchi identity, bosonic equations of motion, and supersymmetry variations of the gravitino. The general class includes an $\cN=2$ AdS$_4$ solution dual to the conformal, low-energy physics phase corresponding to a mass deformation of the M2-brane field theory. A consistent truncation recently constructed on this particular geometry is recovered from our formalism.

\end{quote}

\vfill

\end{titlepage}

\tableofcontents



\section{Introduction}


For every solution of string or M-theory of the form $\textrm{AdS}_D \times Y$, with the product generically warped, supported by fluxes and preserving $\cN$ supersymmetries, a consistent truncation should exist on the internal Riemannian manifold $Y$ down to the $\cN$--extended pure gauged supergravity in $D$ dimensions \cite{Gauntlett:2007ma} (see also \cite{Duff:1985jd,Pope:1987ad}) . Classes of supersymmetric solutions of this type are completely characterised by a $G$-structure \cite{Gauntlett:2002sc} on $Y$, whose $G$--invariant forms are constructed as bilinears of the $\cN$ preserved Killing spinors. These bilinears completely determine the entire class of $\textrm{AdS}_D \times Y$ solutions: the metric on $Y$, the warping factor  and the internal fluxes. By construction, the $G$-structure determines this class of solutions with an AdS$_D$ factor and, in principle, only this class of solutions.

The class of consistent truncations discussed in \cite{Gauntlett:2007ma} turns out to be characterised by the exact same $G$-structure. The higher-dimensional geometry is now of the schematic form $M_D \rtimes \hat{Y}$, with the same warping but with the external $\textrm{AdS}_D$ metric replaced with the metric $g_D$ on a $D$-dimensional spacetime $M_D$. Generically, the geometry on $Y$ gets deformed by the $D$-dimensional scalars if present, and fibred over $M_D$ by the $D$-dimensional vectors, if also present. We have denoted these deformations by $\hat{Y}$. The string or M-theory fluxes also get new contributions containing the $D$-dimensional supergravity fields wedged with forms constructed from those that define the $G$-structure on $Y$. In any case, the latter still governs all these deformations and, thus, completely determines the larger class of string/M-theory solutions $M_D \rtimes \hat{Y}$.

From the point of view of the consistent truncation, the original higher-dimensional $\textrm{AdS}_D \times Y$ solution acquires an alternate interpretation. It may be regarded as the uplift of the vacuum solution of $D$-dimensional gauged supergravity, attained by setting $g_D$ equal to the AdS$_D$ metric and by turning off all other gauged supergravity fields. By the consistency of the truncation, however, {\it any} solution of the $D$-dimensional supergravity, not only the vacuum, must give rise to a higher dimensional solution $M_D \rtimes \hat{Y}$ supported by deformed fluxes, which is completely specified by the same $G$-structure than the original, undeformed solution $\textrm{AdS}_D \times Y$. From this perspective, it is striking that a $G$-structure whose original role was simply to describe a class of supersymmetric $\textrm{AdS}_D \times Y$ solutions with an AdS$_D$ factor, turns out to account, for free, for a much larger (in fact, infinite) class of solutions $M_D \rtimes \hat{Y}$, supersymmetric or otherwise. The $G$-structure was not originally designed to do this, but it nevertheless does.

A free lunch is thus available. It must be emphasised that one still needs to pay the price of finding the deformed geometry and fluxes out of the original $G$-structure, that is, the consistent Kaluza-Klein (KK) truncation ansatz. But once that is done, a powerful solution generating technique is available that is governed by the geometry of only a particular solution within a more general class. In this paper, we will focus on consistent truncations of $D=11$ supergravity \cite{Cremmer:1978km} down to minimal $D=4$ $\cN=2$ gauged supergravity \cite{Fradkin:1976xz,Freedman:1976aw}. Two such classes of consistent truncations were constructed in \cite{Gauntlett:2007ma} (see also \cite{Pope:1985jg}), respectively associated to two classes of $\textrm{AdS}_4 \times Y_7$ M-theory solutions with $\cN=2$ supersymmetry. The first class is of the Freund-Rubin direct product type \cite{Freund:1980xh}, with purely electric four-form flux along AdS$_4$ and with $Y_7$ equipped with a Sasaki-Einstein structure, see \cite{Duff:1984hn}. This class of solutions arise as the near horizon geometry of M2-branes probing a Calabi-Yau four-fold singularity and is the most general class of $\cN=2$ AdS$_4$ solutions with purely electric four-form flux. The second class of solutions were discussed in \cite{Gauntlett:2006ux}, and is the most general $\cN=2$ AdS$_4$ class of M-theory solutions with purely magnetic flux. This class of geometries is related to M5-branes wrapping special lagrangian (SLAG) three-cycles inside $Y_7$. 

In \cite{Gabella:2012rc}, the most general class of $\cN=2$ AdS$_4$ M-theory solutions was constructed using $G$-structure techniques. This class of solutions \cite{Gabella:2012rc} is supported by general four-form flux, and encompasses the purely electric \cite{Duff:1984hn} and purely magnetic \cite{Gauntlett:2006ux} cases as particular limits. In this paper, we will construct the predicted \cite{Gauntlett:2007ma} consistent truncation of $D=11$ supergravity to $D=4$ $\cN=2$ gauged supergravity on the general $\cN=2$ geometries of \cite{Gabella:2012rc}. We will build a truncation ansatz for the bosonic fields and show its consistency at the level of the bosonic equations of motion. We will also show that the supersymmetry variations of the $D=11$ gravitino truncate consistently into their $D=4$ counterparts. Thus, our results show that any (bosonic) solution of $D=4$ $\cN=2$ minimal gauged supergravity can be consistently uplifted to $D=11$ on the seven-dimensional class of geometries of \cite{Gabella:2012rc}, and that the resulting $D=11$ configuration will preserve any supersymmetries, up to $\cN=2$, that the $D=4$ configuration might have. Other consistent truncations to pure gauged supergravities in various dimensions on different $G$-structure geometries have been constructed in \cite{Tsikas:1986rx,Buchel:2006gb,Gauntlett:2006ai,Gauntlett:2007sm,Passias:2015gya,Malek:2017njj,Hong:2018amk,Malek:2018zcz,Liu:2019cea,Cassani:2019vcl}.


\section{The background geometry} \label{sec: undefSol}


We start by reviewing, for later reference, the class of background geometries of \cite{Gabella:2012rc}. These correspond to warped products $\textrm{AdS}_4 \times Y_7$ with $D=11$ metric and four-form
\begin{equation}	\label{eq: vacuum}
	g_{11} = e^{2\Delta}\left(g_{\text{AdS}_4}+g_7\right)	\; , \qquad 
	G_\4 =m\, \vol(\text{AdS}_4)+F_\4 \; ,
\end{equation}
where $m$ is a constant and the function $e^{2\Delta}$, the Riemannian metric $g_7$ and the four-form $F_\4$ are all defined on the internal manifold $Y_7$. We follow \cite{Gabella:2012rc} in defining $g_{\text{AdS}_4}$ to be of radius $L_{\text{AdS}_4}=\frac12$ so that its Ricci tensor is $-12$ times the metric. In (\ref{eq: vacuum}), $\vol(\text{AdS}_4)$ is the volume form of $g_{\text{AdS}_4}$. Two linearly independent Dirac spinors $\chi_i$, $i=1,2$, are defined on $Y_7$, which are subject to the constraints
\begin{equation}	\label{eq: KillingSpinorEq}
	\begin{aligned}
		\frac12\partial_n\Delta\g^n\chi_i-\frac{ime^{-3\Delta}}{6}\chi_i+\frac{e^{-3\Delta}}{288}F_{bcde}\g^{bcde}\chi_i+\chi_i^c=0\;,		\\
		\grad_m\chi_i+\frac{ime^{-3\Delta}}{4}\g_m\chi_i-\frac{e^{-3\Delta}}{24}\g^{cde}F_{mcde}\chi_i-\g_m\chi_i^c=0\; ,				
	\end{aligned}
\end{equation}
imposed by the requirement that the $D=11$ configuration (\ref{eq: vacuum}) preserves $\cN=2$ supersymmetries. Indices $a,b,c , \ldots = 4 , \ldots , 10$ and $m,n,p , \ldots = 4 , \ldots , 10$ respectively are $M_7$ global and local indices, $\gamma_a$ and $\gamma_m$ respectively denote the seven-dimensional Dirac matrices and their contraction with a local frame, $\grad_m$ is the covariant derivative compatible with $g_7$ acting on spinors, and the superscript $c$ here (and in (\ref{eq: KillingSpinorUndef}) below) stands for charge conjugate with the standard conventions of \cite{Gabella:2012rc}.

A number of bilinears in $\chi_i$ can be constructed that define a local SU(2) structure on $Y_7$. This is ultimately specified by a triplet of orthonormal spinor bilinear one-forms, $E_1$, $E_2$, $E_3$, and two-forms, $J_1$, $J_2$, $J_3$. One of the one-forms, $E_1$, is dual to a Killing vector $\xi$ of $g_7$ that also preserves the four-form flux $F_\4$. This vector thus generates the Reeb-like $\cN=2$ direction. Local coordinates $\psi$, $\tau$ and $\rho$ can be introduced on $Y_7$ so that the Killing vector is $\xi=4\partial_\psi$, and the one-forms become
\begin{equation}	\label{eq: candreibein}
	E_1=\frac14\norm{\xi}(\dd \psi +\mathcal{A})\,,		\qquad
	E_2=\frac{e^{-3\Delta}}{4\sqrt{1-\norm{\xi}^2}}\dd \rho\,,	\qquad
	E_3=\frac6m\,\frac{\rho\norm{\xi}}{4\sqrt{1-\norm{\xi}^2}}(\dd \tau +\mathcal{A})\,,
\end{equation}
where $\norm{\xi}$ is the norm of $\xi$ with respect to $g_7$,
\begin{equation} \label{eq:Normxi}
\norm{\xi}^2 = \frac{e^{-6\Delta}}{36} \, \big( m^2 +36 \rho^2 \big) \; ,
\end{equation}
and $\mathcal{A}$ is a local one-form such that $\mathcal{L}_{\xi}\mathcal{A}=0$ and $i_\xi\mathcal{A}=0$.  

The metric on $M_7$ can now be written as
\begin{equation}	\label{eq: 7metric}
	g_7=g_{\text{SU}(2)}+E^2_1+E^2_2+E^2_3\,,
\end{equation}
with $g_{\text{SU}(2)}$ a metric on the local four-dimensional space where the two-forms $J_I$, $I=1,2,3$, are defined. In terms of a frame on this space\footnote{\label{eq: unwarpedelfbein} We label the $D=11$ frame so that  $ g_4+g_7=-e^0\otimes e^0+\sum_{i=1}^{10}e^i\otimes e^i$, with $e^0,\dots,e^3$ associated to AdS$_4$, $e^4,\dots,e^7$ to $g_{\text{SU}(2)}$, and $e^8=E_1$, $e^9=E_2$, $e^{10}=E_3$.}, these take on the canonical expressions
\begin{equation}
		J_3 =e^{45}+e^{67}	\; , \qquad 
		\Omega = J_1+iJ_2 =(e^4+ie^5)\wedge(e^6+ie^7)\,.
\end{equation}
In particular, $J_I$ are self-dual with respect to the Hodge star associated to $g_{\text{SU}(2)}$ and obey $J_I\wedge J_J = 2\, \vol (g_{\text{SU}(2)}) \,\delta_{IJ}$.

Finally, the SU(2)-structure forms satisfy the following torsion conditions 
%
	\begin{align}
		e^{-3\Delta}\dd \left[\norm{\xi}^{-1}\left(\frac m6E_1+e^{3\Delta}\vert S\vert\sqrt{1-\norm{\xi}^2}E_3\right)\right]&=2\left(J_3-\norm{\xi} E_2 \wedge E_3 \right)\,,	\label{eq: TC1}\\[6pt]
		\dd \left(\norm{\xi}^2e^{9\Delta}J_2\wedge E_2\right)-e^{3\Delta}\vert S\vert \dd \left(\norm{\xi}e^{6\Delta}\vert S\vert^{-1}J_1\wedge E_3\right)&=0\,,		\label{eq: TC2}\\[6pt]
		\dd \left(e^{6\Delta}J_1\wedge E_2\right)+e^{3\Delta}\vert S\vert \dd \left(\norm{\xi}e^{3\Delta}\vert S\vert^{-1}J_2\wedge E_3\right)&=0\,,				\label{eq: TC3}
	\end{align}
%
where $S \equiv \rho e^{-3\Delta} e^{i (\psi - \tau)}$ is a zero-form bilinear. These determine the internal four-form as
\begin{equation}	\label{eq: undefFlux}
	F_\4=\frac1{\norm{\xi}}E_1\wedge \dd \left(e^{3\Delta}\sqrt{1-\norm{\xi}^2}J_1\right)-m\frac{\sqrt{1-\norm{\xi}^2}}{\norm{\xi}}J_1\wedge E_2\wedge E_3 \,,
\end{equation}
and the differential of the one-form ${\cal A}$ as 
\begin{equation} \label{eq:dcalA}
\dd \mathcal{A} =  \frac{4m e^{-3\Delta}}{3\norm{\xi}^2}\left[J_3 + \left(3\norm{\xi}-\frac{4}{\norm{\xi}}\right)E_2\wedge E_3\right] \; .
\end{equation}

The supersymmetric configuration (\ref{eq: vacuum}) with (\ref{eq: candreibein})--(\ref{eq:dcalA}) solves the Bianchi identities and equations of motion (\ref{eq: Meoms}) of $D=11$ supergravity \cite{Gabella:2012rc} (see also \cite{Gauntlett:2002fz} for general statements about supersymmetry implying the equations of motion for $G$-structure solutions). In particular, it is straightforward to check that the four-form (\ref{eq: undefFlux}) is closed, using the differential relations (\ref{eq: TC1})--(\ref{eq: TC3}). In fact, the two distinct contributions to the four-form can be checked to be separately closed. 


\section{Consistent truncation} \label{sec: KK}


We now turn to construct the consistent truncation of $D=11$ supergravity \cite{Cremmer:1978km} on the seven-dimensional geometries of \cite{Gabella:2012rc} that were reviewed in section \ref{sec: undefSol}, down to minimal $D=4$ $\cN=2$ gauged supergravity \cite{Fradkin:1976xz,Freedman:1976aw}. Our conventions for these theories are specified in appendix \ref{sec:Conventions}.

For that purpose, we propose the following KK ansatz:
\begin{equation}	\label{eq:ansatz}
		g_{11} =e^{2\Delta}(g_4+\hat{g}_7)	\; , \qquad 
		G_\4 =m\, \vol_4+\hF_\4-\a\wedge\,g  \bar{F} -\beta\wedge\,g\star_4 \bar{F} \; .
\end{equation}
The metric $g_4$ is a general $D=4$ metric and $\vol_4$ its corresponding volume form. Hats over $\hat{g}_7$ and $\hF_\4$ have been employed to signify a shift of the Reeb direction $\xi$ by the $D=4$ graviphoton $\bar{A}$. The latter is only expected to enter the KK ansatz by gauging these shifts, as in {\it e.g.} \cite{Gauntlett:2007ma}. This motivates, from (\ref{eq: candreibein}), the definition
\begin{equation} \label{eq:E1shifted}
\hat{E}_1=\frac14\norm{\xi}(\dd \psi +\mathcal{A}-g\bar{A}) \; .
\end{equation}
Accordingly we have, from (\ref{eq: 7metric}) and (\ref{eq: undefFlux}),
\begin{equation} \label{Hattedg7F4}
	\begin{aligned}
		\hat{g}_7&=g_{SU(2)}+\hat{E}_1^2+E_2^2+E_3^2 \quad ,	\\[.5em]
		\hF_\4&=\frac1{\norm{\xi}}\hat{E}_1\wedge \dd \left(e^{3\Delta}\sqrt{1-\norm{\xi}^2}J_1\right)-m\frac{\sqrt{1-\norm{\xi}^2}}{\norm{\xi}}J_1\wedge E_2\wedge E_3  \; .
	\end{aligned}
\end{equation}
The graviphoton also enters the KK ansatz (\ref{eq:ansatz}) through its field strength $ \bar{F} =\dd\bar{A}$ and through the Hodge dual of the latter with respect to the four-dimensional metric $g_4$. The constant $g$ that appears in (\ref{eq:ansatz}) and (\ref{eq:E1shifted}) is the gauge coupling of the $D=4$ supergravity. Finally, $\alpha$ and $\beta$ are two-forms on the internal seven-dimensional manifold to be determined.

When $g_4$ is set equal to the AdS$_4$ metric and the graviphoton is turned off, $\bar{A} = 0$, $\bar{F} = 0$, the $D=11$ configuration (\ref{eq:ansatz}) reduces to the $\cN=2$ class of solutions of \cite{Gabella:2012rc}. In this case, the two-forms $\alpha$, $\beta$ drop out from the picture and do not play any role in the background geometry. More generally, though, the full configuration (\ref{eq:ansatz}) with general $D=4$ fields $g_4$, $\bar{A}$ subject to the field equations of $D=4$ $\cN=2$ minimal supergravity, can still be forced to obey the field equations of $D=11$ supergravity for suitable $\alpha$ and $\beta$. The strategy is to substitute (\ref{eq:ansatz}) into the $D=11$ field equations treating the linear, $\bar{F}$, $\star_4 \bar{F}$, and quadratic,  $\bar{F} \wedge  \bar{F}$, $\star_4 \bar{F} \wedge  \bar{F}$, combinations of the $D=4$ graviphoton field strength as independent quantities. Upon imposing the $D=4$ field equations, a number of differential and algebraic equations for $\alpha$ and $\beta$ are produced. Proposing a suitable ansatz for these two-forms in terms of the SU(2)--structure forms and using the torsion conditions (\ref{eq: TC1})--(\ref{eq: TC3}), we can solve this system of equations and, thus, find the explicit consistent KK reduction.

Let us summarise, along these lines, the system of equations that $\alpha$ and $\beta$ must obey for the truncation ansatz to be consistent. Further details on the consistency proof are relegated to appendix \ref{sec:KKproofEoms}. In our conventions, the $D=11$ and $D=4$ field equations take on the form (\ref{eq: Meoms}) and (\ref{minimalN=2eoms}).  It is convenient to introduce the two-forms $\ta$, $\tb$ containing the contributions to $\alpha$, $\beta$ with no legs along the gauged $E_1$ direction (see the appendix). Imposing the Bianchi identity for the undeformed four-form in (\ref{eq: vacuum}), and the Bianchi and Maxwell equation for the $D=4$ graviphoton, the Bianchi identity of the deformed four-form in (\ref{eq:ansatz}) is satisfied provided the unknown forms obey the following constraints:
\begin{alignat}{2}	\label{eq: BianchitwoformsText}
	 \bar{F} \wedge  \bar{F} :&\quad	i_{\xi}\a=0 \; ,							&& \bar{F} :\quad \frac 14i_\xi F_\4+\dd\ta=0	\; ,	\notag\\
	\star_4 \bar{F} \wedge  \bar{F} :&\quad i_{\xi}\b=0 \; ,	\hspace{3cm}		\star_4&& \bar{F} :\quad \dd\tb=0	\; .
\end{alignat}
These expressions arise in the $D=11$ five-form $\dd G_\4 = 0 $ wedged with the indicated $D=4$ graviphoton contributions, and must be enforced to vanish separately for arbitrary $\bar{F}$. The constraints coming from the quadratic graviphoton contributions imply $\alpha = \ta$, $\beta = \tb$. We will make use of these relations in the sequel to simplify the resulting expressions. 

Proceeding similarly, we find the constraints imposed on $\alpha$ and $\beta$ by the equation of motion for the $D=11$ four-form. Assuming, again, that the undeformed four-form (\ref{eq: vacuum}) satisfies the equation of motion and imposing the Bianchi and equation of motion for $\bar{F}$, the equation of motion for $G_\4$ in (\ref{eq:ansatz}) is satisfied provided the following relations hold:
%
%
	\begin{align}	
		 \bar{F} \wedge  \bar{F} :
			&	\quad	\frac14e^{3\Delta}i_\xi\star_7\b+\frac12(\b\wedge\b-\a\wedge\a) = 0 \;,	\nonumber \\
		\star_4 \bar{F} \wedge  \bar{F} :
			&	\quad	\frac14e^{3\Delta}i_\xi\star_7\a+\a\wedge\b = 0  \;,					\nonumber \\
		 \bar{F} :
			&	\quad	\frac m8\norm{\xi}e^{3\Delta}J_3\wedge J_3\wedge E_2\wedge E_3-\frac14e^{3\Delta}\dd \mathcal{A}\wedge i_{\xi}\star_7\b	\notag\\
			&	\qquad\quad		+\frac14\he\wedge \dd\, (e^{3\Delta}i_\xi\star_7\b)+\a\wedge\hF_\4  = 0   \;, \nonumber \\
				 \label{eq: EomswoformsText}
		\star_4 \bar{F} :
			&	\quad	\frac14e^{3\Delta}\dd \mathcal{A}\wedge i_{\xi}\star_7\a-\frac14\he\wedge \dd\, (e^{3\Delta}i_\xi\star_7\a)+\b\wedge\hF_\4 = 0  \; ,
	\end{align}
%
with $\hat e$ defined below (\ref{eq:ehatEqs}). We have again indicated the linear or quadratic graviphoton combinations with which these expressions appear wedged in the $D=11$ eight-form equation of motion for $G_\4$. 

Finally, we turn to the evaluation of the $D=11$ Einstein equation on the configuration (\ref{eq:ansatz}). Combining the Ricci tensor (\ref{eq:Ricci}) and the r.h.s.~(\ref{eq:SEt}) of the Einstein equation as given in (\ref{eq: Meoms}), this yields the following three equations,
{\setlength\arraycolsep{0pt}
\begin{eqnarray} \label{eq:EinsteinExternalText}
&& \text{Ric}_{\a\b}-\frac{g^2}{32}\norm{\xi}^2\bar{F}_{\a\g}\bar{F}_\b{}^\g-9(\partial_a\Delta\partial^a\Delta+\nabla_a\nabla^a\Delta)\eta_{\a\b}				\notag	\\
			&& \quad =-e^{-6\Delta}\left\{\frac13m^2\eta_{\a\b}-\frac{g^2}4(\a^2+\b^2)\bar{F}_{\a\g}\bar{F}_\b{}^\g+\frac{g^2}{24}\eta_{\a\b}\bar{F}^2(\a^2+2\b^2)\right.	\notag	\\
			&&	\quad  \quad \qquad\qquad+\left.\frac{g^2}4\bar{F}_{\g(\a}?\epsilon_\b)^\g\mu\nu?\bar{F}_{\mu\nu}\a_{cd}\b^{cd}
			+\frac{g^2}{24}\eta_{\a\b}\epsilon_{\mu\nu\rho\sigma}\bar{F}^{\mu\nu}\bar{F}^{\rho\sigma}\a_{cd}\b^{cd}		\right\}\;, \\[10pt]
&& \label{eq:EinsteinMixedText}	\frac g8\norm{\xi}\delta_{8b}\nabla_\g \bar{F}_\a{}^\g=0 \; ,
\\[10pt]
\label{eq:EinsteinInternalText}
&& 	\text{Ric}_{ab}	+\frac{g^2}{64}\norm{\xi}^2\delta_{8a}\delta_{8b} \bar{F}_{\g\d}\bar{F}^{\g\d}+9[\partial_a\Delta\partial_b\Delta-\nabla_a\nabla_b\Delta
			-(\partial_c\Delta\partial^c\Delta-\nabla_c\nabla^c\Delta)		\notag	\\
			&& \quad =e^{-6\Delta}\left\{\frac12\left[F_{acde}?F_b^cde?-\frac1{12}\eta_{ab}F^2\right]+\frac{g^2}{24}\bar{F}^2\left[6(\a_{ac}?\a_b^c?-\b_{ac}?\b_b^c?)
			-\eta_{ab}(\a^2-\b^2)\right]\right.							\notag	\\
			&& \left. \quad \qquad\qquad  +\frac{g^2}{24}\epsilon_{\mu\nu\rho\sigma}\bar{F}^{\mu\nu}\bar{F}^{\rho\sigma}\left[3(\a_{ac}?\b_b^c?+\b_{ac}?\a_b^c?)-\eta_{ab}\a_{cd}\b^{cd}\right]
			+\frac12m^2\eta_{ab}\right\}\; ,
\end{eqnarray}
}with $\alpha = 0 , \ldots , 3$ and $a = 4 , \ldots , 10$ external and internal tangent space indices related to the frame specified in footnote \ref{eq: unwarpedelfbein}. Also, $\a^2=\a_{ab}\a^{ab}$ and similarly for $\b^2$, $F^2$ and $\bar{F}^2$. In (\ref{eq:EinsteinExternalText}) and (\ref{eq:EinsteinInternalText}), $\text{Ric}_{\a\b}$ and $\text{Ric}_{ab}$ are the Ricci tensors of $g_4$ and the undeformed $g_7$ metric. Expectedly, the only non-trivial mixed components, (\ref{eq:EinsteinMixedText}), of the Einstein equations arise in the direction (the 8-th in the notation of footnote \ref{eq: unwarpedelfbein}) that is gauged. The resulting equation is automatically satisfied on the graviphoton's Maxwell equation (the second equation in (\ref{minimalN=2eoms})).

For suitably chosen $\alpha$ and $\beta$ in terms of background SU(2)-structure forms, equations (\ref{eq: BianchitwoformsText})--(\ref{eq:EinsteinInternalText}) must be satisfied identically, and equation (\ref{eq:EinsteinExternalText}) must reduce to the $D=4$ Einstein equation. As shown in appendix \ref{sec:KKproofEoms}, all these requirements are satisfied by setting
\begin{equation} \label{eq:KKFormCoefs}
\a =-\frac14e^{3\Delta}\sqrt{1-\norm{\xi}^2}J_1 \; , \qquad
\b =-\frac14e^{3\Delta}\left(J_3-\norm{\xi}E_2\wedge E_3\right) \; .
\end{equation}
The KK ansatz (\ref{eq:ansatz}) is thus consistent, at the level of the bosonic field equations, when the two-form coefficients $\alpha$, $\beta$ are taken as in (\ref{eq:KKFormCoefs}). Furthermore, consistency can be extended to include the fermions, as we now turn to discuss at the level of the supersymmetry variations of the gravitino. See appendix \ref{sec:KKproofSusy} for further details. 

We start by factorising the Majorana spinor parameter $\epsilon$ in terms of two $D=4$ Weyl spinor parameters of positive chirality $\psi_i^+$, $i=1,2$, and the Dirac spinors $\chi_i$ on the undeformed internal seven-dimensional space, formally as in \cite{Gabella:2012rc}, 
\begin{equation}	\label{eq: KillingSpinorUndef}
	\epsilon =\psi_i^+\otimes e^{\Delta/2}\chi_i+(\psi_i^+)^c\otimes e^{\Delta/2}\chi_i^c \; .
\end{equation}
The sum here extends over $i=1,2$, and the factors of $e^{\Delta/2}$ have been chosen, as in \cite{Gabella:2012rc}, for convenience. The only difference with respect to \cite{Gabella:2012rc} is that the parameters $\psi_i^+$ in (\ref{eq: KillingSpinorUndef}) are no longer subject to the AdS$_4$ Killing spinor equations. Next, we plug the KK ansatz (\ref{eq:ansatz}) with (\ref{eq:KKFormCoefs}) into the $D=11$ gravitino variation (\ref{eq: susyvar}), written in the basis (\ref{eq: CliffSplitting}) for the $D=11$ Dirac matrices in terms of their four-, $\rho_\alpha$, and seven-dimensional, $\gamma_a$, counterparts. Then, we address the internal and external gravitino variations separately.

A long calculation, summarised in appendix \ref{sec:KKproofSusy}, shows that the internal gravitino variations vanish identically provided the following projections,
\begin{equation}		\label{eq: cond8}
	\begin{aligned}
		\left[\norm{\xi}(3\g^8+i\g^{910})+\sqrt{1-\norm{\xi}^2}(\g^{46}-\g^{57})-i(\g^{45}+\g^{67})\right]\chi_i=0\;,
	\end{aligned}
\end{equation}
and
\begin{eqnarray}	\label{eq: proj3}
&	(\g^{46}+\g^{57})\chi_i=0,
	\qquad
	(\g^{45}-\g^{67})\chi_i=0\;, \nonumber \\[5pt]
&		\left[-\sqrt{1-\norm{\xi}^2}\g^{46}+i(\g^{45}+\norm{\xi}\g^{910})\right]\chi_i=0 \; ,
\end{eqnarray}
are imposed on the internal spinors $\chi_i$. These projections, however, add nothing new: they follow from the undeformed Killing spinor equations (\ref{eq: KillingSpinorEq}) of the undeformed geometry. This is best seen by sandwiching (\ref{eq: cond8}), (\ref{eq: proj3}) with the conjugate spinors $\bar{\chi}_j$: the resulting constraints are identically satisfied by the spinor bilinears that defined the undeformed SU(2)--structure. The internal gravitino variations are thus automatically satisfied for the general class of solutions (\ref{eq:ansatz}), using only the restrictions that characterise the AdS$_4$ solutions (\ref{eq: vacuum}).

The calculation of the external gravitino variations proceeds similarly. Together with (\ref{eq: cond8}), (\ref{eq: proj3}), the following projection must be imposed:
\begin{equation} \label{eq:PrjGravExt}
	i\g^{45}\chi_i=-\e_{ij}\chi_j^c\;.
\end{equation}
This, like (\ref{eq: cond8}), (\ref{eq: proj3}), is still compatible with the original Killing spinor equations (\ref{eq: KillingSpinorEq}) of the undeformed geometry, as argued in appendix \ref{sec:KKproofSusy}, and does not reduce the amount of supersymmetry or constrain the undeformed geometry further. The calculation allows one to read off the consistent embedding of the $D=4$ $\cN=2$ gravitini $\psi _{i\mu}^+$, $i=1,2$, into its $D=11$ counterpart $\Psi _{M}$, for $M= \mu$:
\begin{equation} \label{eq:KKGravitino}
	\Psi_\mu=\psi^+_{i\mu}\otimes e^{\Delta/2}\chi_i +(\psi^+_{i\mu})^c\otimes e^{\Delta/2}\chi^c_i \; ,
\end{equation}
with sum over $i$. Using (\ref{eq:KKGravitino}), the external components of the $D=11$ gravitino variation (\ref{eq: Meoms}) finally reduce to their $D=4$ $\cN=2$ counterparts, (\ref{eq:4dGravitinoSusy}). 

To summarise, any solution of minimal $D=4$ $\cN=2$ gauged supergravity gives rise to a class of solutions of $D=11$ supergravity of the form 
{\setlength\arraycolsep{2pt}
\begin{eqnarray}	\label{eq:ansatzSolved}
&&		g_{11} =e^{2\Delta}(g_4+\hat{g}_7)	\; ,  \\[5pt]
&& 		G_\4 =m \, \vol_4+\hF_\4 + \frac14e^{3\Delta}\sqrt{1-\norm{\xi}^2}J_1 \wedge\,g  \bar{F} + \frac14e^{3\Delta}\left(J_3-\norm{\xi}E_2\wedge E_3\right) \wedge\,g\star_4 \bar{F} \; , \nonumber
\end{eqnarray}
}with $\hat{g}_7$, $\hF_\4$ defined in (\ref{Hattedg7F4}), upon uplift on the class of seven-dimensional geometries \cite{Gabella:2012rc} reviewed in section \ref{sec: undefSol}. The uplift preserves supersymmetry if originally present in $D=4$. The general class of solutions (\ref{eq:ansatzSolved}) is completely specified by the $D=4$ supergravity fields and the same SU(2)-structure that characterises the background AdS$_4$ class of solutions (\ref{eq: vacuum}) of \cite{Gabella:2012rc}. The free lunch promised by \cite{Gauntlett:2007ma} is now served.


\section{Discussion} 	\label{sec: KillingSol}


It is interesting to determine how our KK truncation ansatz adapts itself to various particular cases of the general geometries of \cite{Gabella:2012rc}. In the purely magnetic flux case, the geometries \cite{Gabella:2012rc} reduce, by apropriately taking the $m=0$ limit, to the $\cN=2$ class of geometries describing M5-branes wrapped on internal SLAG 3-cycles described in \cite{Gauntlett:2006ux}. Accordingly, our consistent truncation reduces to the one considered in section 3 of \cite{Gauntlett:2007ma}. The purely electric, Freund-Rubin class of solutions with Sasaki-Einstein internal space \cite{Duff:1984hn} is not directly obtainable from the generic class that we have been using since, as the authors \cite{Gabella:2012rc} discuss, this geometry is attained for a different choice of internal spinors $\chi_i$. In any case, a consistent truncation to $D=4$ $\cN=2$ supergravity can be also obtained in this case \cite{Gauntlett:2007ma,Pope:1985jg}.

More interestingly, a subcase of the general class of configurations of \cite{Gabella:2012rc} was also studied in that reference, where the vector $\partial_\tau$ along the coordinate $\tau$ becomes an isometry of the internal metric $g_7$. This vector can never become a symmetry of $F_\4$, though, unlike the Reeb vector $\xi = 4 \partial_\psi$, which preserves the entire $D=11$ configuration. Let us particularise our general consistent truncation of section \ref{sec: KK} to this concrete class of solutions. Following \cite{Gabella:2012rc}, we rescale the coordinate $\rho$ by a constant factor as $r \equiv \frac6m \rho$, for convenience, and introduce a function $f(r)$ such that
\begin{equation}
J_I=\frac m{24}e^{-3\Delta} f(r)\,\mathbb{J}_I \; , \; I= 1,2,3 \; , \qquad
(1+r^2)(\dd \tau +\mathcal{A})=f(r)(\dd \tau +A_{\textrm{KE}}) \; ,
\end{equation}
where the one-form $A_{\textrm{KE}}$ and the triplet of two-forms $\mathbb{J}_I$ are $r$--independent and defined on the four-dimensional space with metric $g_{\text{SU}(2)}$. The latter becomes, up to an overall $r$-dependent factor (see (\ref{eq: CPWmetricuplift})), a K\"ahler-Einstein metric $g_{\textrm{KE}}$ with canonical normalisation $\text{Ric}_{\textrm{KE}}=6\,g_{\textrm{KE}}$. The torsion conditions \eqref{eq: TC1}--\eqref{eq: TC3} reduce to
\begin{equation}	\label{eq: KillingFormRelations}
	\dd A_{\textrm{KE}}=2 \, \mathbb{J}_3\,,		\qquad\qquad	
	\dd\,(\mathbb{J}_1+i\mathbb{J}_2)=3 i (\mathbb{J}_1+i\mathbb{J}_2)\wedge (\dd\tau+A_{\textrm{KE}})\,,	
\end{equation}
together with the following set of ordinary differential equations (ODEs) for $f(r)$,
\begin{equation}	\label{eq: KillingODE}
		f'=-\frac12r\, \Omega^2\,  f\,,	\qquad\qquad
		\frac{(r\Omega'-r^2\Omega^3)f}{\sqrt{1+(1+r^2)\Omega^2}}=-3 \, ,
\end{equation}
where a prime denotes derivative with respect to $r$. The function $\Omega(r)$ equals $\alpha(r)$ in \cite{Gabella:2012rc} and is introduced, for convenience, as a substitute of the warp factor. The latter can be reobtained as $e^{6\Delta} = \left( \frac{m}{6} \right)^2 \big( 1 + r^2 + \Omega^{-2}$). The first equation in (\ref{eq: KillingFormRelations}) signals the two-form $\mathbb{J}_3$ as the K\"ahler-Einstein form and $A_{\textrm{KE}}$ as a potential for it. Finally, the one-forms \eqref{eq: candreibein} become, using  (\ref{eq:Normxi}), 
\begin{equation}	\label{eq: killingdreibein}
	\begin{aligned}
	E_1	&=\frac{\Omega\sqrt{1+r^2}}{4\sqrt{1+(1+r^2)\Omega^2}}\left[\dd \psi -\dd\tau+\frac{f}{1+r^2}(\dd\tau+A_{\textrm{KE}})\right]	\,,\\[.7em]
	E_2	&=\frac14\Omega\, \dd r 					\,,\hspace{2.6cm}
	E_3	=\frac14\frac{r\,\Omega\, f}{\sqrt{1+r^2}}(\dd \tau +A_{\textrm{KE}})\,.
	\end{aligned}
\end{equation}

Bringing these definitions to the general consistent truncation formulae of section \ref{sec: KK}, we can obtain the consistent truncation corresponding to the subclass of geometries where $\partial_\tau$ is Killing. After some calculation, we find that the KK ansatz becomes\footnote{When the $D=4$ supergravity fields are turned off, the metric (\ref{eq: CPWmetricuplift}) agrees, up to a straightforward redefinition of $\psi$, with (4.13) of \cite{Gabella:2012rc}. However, the background four-form (\ref{eq: CPW4formuplift}) seems to disagree with their (4.14).}
\begin{equation}	\label{eq: CPWmetricuplift}
	\begin{aligned}
		g_{11}&=e^{2\Delta}\left\{g_4+\frac{\Omega\,f}{4\sqrt{1+(1+r^2)\Omega^2}}\, g_{\textrm{KE}}+\frac{\Omega^2}{16}\left[\dd r ^2+\frac{r^2f^2}{1+r^2}(\dd \tau +A_{\textrm{KE}})^2	\right.\right.\\
		&\quad\qquad\left.+\left.\frac{1+r^2}{1+(1+r^2)\Omega^2}\left(\text{D} \psi-\dd\tau +\frac{f}{1+r^2}\left(\dd \tau +A_{\textrm{KE}}\right)\right)^2\right]\right\}\, ,
	\end{aligned}
\end{equation}
\begin{equation}	\label{eq: CPW4formuplift}
	\begin{aligned}
		\hF_\4&=h_1(r)(\text{D} \psi -\dd\tau)\wedge \dd r \wedge\mathbb{J}_1+h_2(r)(\text{D} \psi-\dd\tau)\wedge (\dd \tau +A_{\textrm{KE}})\wedge \mathbb{J}_2		\\
		&\quad+h_3(r)(\dd \tau +A_{\textrm{KE}})\wedge \dd r \wedge \mathbb{J}_1-\a\wedge g\bar{F}-\b\wedge g\star_4\bar{F}\, ,
	\end{aligned}
\end{equation}
where we have defined the following shorthand functions of $r$
\begin{equation}	\label{eq: haches}
	\begin{aligned}
		h_1(r)	=&\frac{m^2}{3^2\cdot2^6}\left(\Omega^{-1}e^{-3\Delta}f\right)'\,,			\qquad \qquad 
		h_2(r)	=-\frac{m^2}{3\cdot2^6}\left(\Omega^{-1}e^{-3\Delta}f\right)\,,				\\[4pt]
		&h_3(r)	=\frac{m^2}{3^2\cdot2^7}\frac{f}{1+r^2}\left[2\left(\Omega^{-1}e^{-3\Delta}f\right)'-3r\,\Omega^2\left(\Omega^{-1}e^{-3\Delta}f\right)\right]\, .
	\end{aligned}
\end{equation}
The $D=11$ metric $g_{11}$ depends on the $D=4$ metric $g_4$, explicitly, and on the $D=4$  graviphoton $\bar{A}$ through the gauge covariant derivative $\textrm{D} \psi=\dd \psi-g\bar{A}$. The latter also enters the $D=11$ four-form (\ref{eq: CPW4formuplift}) through its field strength $\bar{F}$ and its Hodge dual. These contributions are wedged with internal forms $\alpha$, $\beta$ which now read, from (\ref{eq:KKFormCoefs}),
\begin{equation}	\label{eq: CPWtwoforms}
	\a=-\frac{m^2}{576}\left(\Omega^{-1}e^{-3\Delta}f\right)\mathbb{J}_1\, , \qquad
	\b=-\frac{mf}{96}\left[\mathbb{J}_3-\frac14r\,\Omega^2 \dd r\wedge(\dd\tau+A_{\textrm{KE}})\right]
	\, .
\end{equation}

Explicit instances in this subclass of geometries are obtained for each solution $f(r)$ of the ODE system (\ref{eq: KillingODE}). Then, (\ref{eq: CPWmetricuplift})--(\ref{eq: CPWtwoforms}) define the corresponding consistent truncation. Two such solutions of (\ref{eq: KillingODE}) were discussed in \cite{Gabella:2012rc}. The first one, analytic, is obtained by setting \cite{Gabella:2012rc}
\begin{equation}	\label{eq: cpwsol}
	f(r)=3\left(2-\frac{r}{\sqrt2}\right),	\qquad
	\Omega(r)=\sqrt{\frac2{2\sqrt2r-r^2}}\;,
\end{equation}
with $r\in[0,\; 2\sqrt2]$. This reproduces the $\cN=2$ AdS$_4$ solution first obtained by Corrado, Pilch and Warner (CPW) \cite{Corrado:2001nv}. Together with Ntokos, we recently obtained a consistent truncation of $D=11$ supergravity on the CPW solution to minimal $D=4$ $\cN=2$ su\-per\-gra\-vi\-ty using other methods \cite{Larios:2019kbw}. Now, we can reproduce that result from these expressions. Using the explicit functions (\ref{eq: cpwsol}), fixing the Freund-Rubin coefficient as $m=\frac8{\sqrt3}\,g^{-3}$, and identifying the internal background geometry quantities here and in \cite{Larios:2019kbw} as
\begin{align}
	r_{\text{here}}=2\sqrt2\sin^2\!\alpha_{\text{there}}&\,,			\qquad
	(\dd \psi -\dd\tau)_{\text{here}}=-2\dd\psi_{\text{there}}'\,,		\qquad
	(\dd \tau +A_{\textrm{KE}})_{\text{here}}=\bm{\eta}'_{\text{ there}}\,,	\notag	\\[.5em]
	&{\mathbb{J}_3}_{\text{here}}=\bm{J}'_{\text{ there}}\,,		\qquad\qquad
	(\mathbb{J}_1+i\mathbb{J}_2)_{\text{here}}=\bm{\Omega}'_{\text{ there}}\,,
\end{align}
the consistent embedding \eqref{eq: CPWmetricuplift}, \eqref{eq: CPW4formuplift} above perfectly matches (3.27), (3.30) of \cite{Larios:2019kbw}. In the latter reference, the consistency of the embedding was guaranteed by construction but, for further reassurance, the Bianchi identities and equation of motion of the $D=11$ four-form field strength were verified to indeed check out. In this paper, we extend the verification of consistency by a re-check of the four-form field equations using $G$-structure technology, and by additional consistency checks at the level of the Einstein equation and the gravitino supersymmetry variations. Incidentally, these provide extra checks on the $\cN=8$ consistent truncations formulae of \cite{Varela:2015ywx}. 

A second, numerical, solution to the ODE system (\ref{eq: KillingODE}) was obtained in \cite{Gabella:2012rc} (see also \cite{Halmagyi:2012ic}). This AdS$_4$ solution was argued \cite{Gabella:2012rc} to dominate holographically the low-energy physics of a relevant deformation of the ABJM \cite{Aharony:2008ug} field theory defined on a stack of planar M2-branes, which is cubic in the adjoint $\cN=2$ chiral fields. Its physical role is thus similar to the CPW solution, which is related to an analogue, quadratic, deformation in the chirals. Associated to this background solution there also exists a consistent truncation to minimal $\cN=2$ supergravity. It is obtained by bringing the corresponding solution $f(r)$ of (\ref{eq: KillingODE}) to (\ref{eq: CPWmetricuplift})--(\ref{eq: CPWtwoforms}).

As a concluding remark, it is interesting to note that our results bring together in $D=11$ the separate classification efforts of \cite{Caldarelli:2003pb,Cacciatori:2004rt} and \cite{Gabella:2012rc}. The supersymmetric solutions of $D=4$ $\cN=2$ minimal gauged supergravity were classified in \cite{Caldarelli:2003pb,Cacciatori:2004rt}. By the consistency of our uplift, any such $D=4$ solution can be fibred over any of the seven-dimensional manifolds of \cite{Gabella:2012rc} to produce, via (\ref{eq:ansatzSolved}), a supersymmetric solution of $D=11$ supergravity.


\section*{Acknowledgements}


We thank Niall Macpherson, Dario Martelli and Achilleas Passias for discussions and correspondence. GL is supported by an FPI-UAM predoctoral fellowship. OV is sup\-por\-ted by the NSF grant PHY-1720364. GL and OV are partially sup\-por\-ted by grants SEV-2016-0597, FPA2015-65480-P and PGC2018-095976-B-C21 from MCIU/AEI/FEDER, UE.

\appendix
\addtocontents{toc}{\protect\setcounter{tocdepth}{1}}


\section{Conventions} \label{sec:Conventions}


We follow the conventions of \cite{Gabella:2012rc} for $D=11$ supergravity \cite{Cremmer:1978km}. The bosonic field content includes the metric $g_{MN}$, $M = 0 , \ldots, 10$, and a three-form potential $A_\3$ with four-form field strength $G_\4 = \dd A_\3$. The bosonic Lagrangian is 
\begin{eqnarray}
{\cal L}_{11} \!\! &=& \!\! R\,  \textrm{vol}_{11} - 
\ft12 \,  G_\4 \wedge \star_{11} G_\4   -\ft16 \, A_\3 \wedge G_\4  \wedge G_\4 \;  , \label{11DLagrangian}
\end{eqnarray}
%
and the field equations
\begin{equation}	\label{eq: Meoms}
	\begin{aligned}
		\dd G_\4&=0\, ,	\\
		\dd\, \star_{11}G_\4+\frac12G_\4\wedge G_\4&=0\, ,		\\
		R_{MN}-\frac1{12}\left[G_{MPQR}G_N{}^{PQR}-\frac1{12}G^2g_{MN}\right]&=0\, .
	\end{aligned}
\end{equation}
The first relation here is the Bianchi identity for $G_\4$, and the other two are the equations of motion that follow from the Lagrangian (\ref{11DLagrangian}). The full action is invariant under local supersymmetry. The gravitino variation reads
\begin{equation}	\label{eq: susyvar}
	\delta_{\e}\Psi_M=\grad_M\e+\frac1{288}\left(\Gamma_M{}^{SPQR}-8\d_M^S\Gamma^{PQR}\right)G_{SPQR}\e=0\,,
\end{equation}
where $\epsilon$ is Majorana and $\Gamma^{A_1 \ldots A_n}$ are the Dirac matrices and their antisymmetrised products. In (\ref{eq: susyvar}), they appear contracted with a local frame. 

The bosonic sector of pure $D=4$ $\cN=2$ supergravity \cite{Fradkin:1976xz,Freedman:1976aw} includes the metric, $\bar{g}_{\mu\nu}$, $\mu = 0 , \ldots 3$, and a gauge field $\bar{A}$, the graviphoton, with field strength $\bar{F} = \dd \bar{A}$. The gauged supergravity has a cosmological constant related to the coupling constant $g$ that couples $\bar{A}$ to the $\cN=2$ gravitini. In our conventions, the bosonic Lagrangian is 
\begin{equation}	\label{eq: minN2sugra}
	\L=\bar{R}\,\overline{\vol}_4-\frac12\bar{F}\wedge\bar{\star}_4\bar{F}+6g^2 \, \overline{\vol}_4 \; ,
\end{equation}
and the field equations
\begin{equation} \label{minimalN=2eoms}
\dd \bar{F} = 0 \; , \qquad 
\dd \bar{\star}_{4} \bar{F} = 0 \; , \qquad 
\bar{R}_{\mu\nu} = -3 g^2 \bar{g}_{\mu\nu} + \tfrac12 \big( \bar{F}_{\mu \sigma} \bar{F}_\nu{}^\sigma -\tfrac14 \bar{g}_{\mu\nu} \, \bar{F}_{\rho \sigma} \bar{F}^{\rho \sigma}  \big) \; .
\end{equation}
Again, the first relation here is the Bianchi identity for $\bar{F}$, and the other two are the equations of motion that follow from the Lagrangian (\ref{eq: minN2sugra}). The theory has two Weyl gravitini, $\psi _{i\mu}^+$. Their variation under supersymmetry is 
\begin{equation}	\label{eq:4dGravitinoSusy}
	\d \psi _{i\mu}^+=\grad_\mu\psi^+_i+\frac {ig}2\e_{ij}\bar{A}_\mu\psi^+_j-\frac g2\bar{\rho}_\mu(\psi^+_i)^c
			+\frac{g^2}{32}\bar{F}_{\d\e}\,\bar{\rho}^{\d\e}\bar{\rho}_\mu\e_{ij}(\psi^+_j)^c\; ,
\end{equation}
for a Weyl spinor parameter $\psi^+_i$ and $\bar{\rho}_\mu$ associated to a local frame for $\bar{g}_4$. 

The $D=11$, $D=4$ and $D=7$ Dirac matrices, $\Gamma_A$, $A= 0 , \ldots , 10$, $\rho_\alpha$, $\alpha = 0 , \ldots, 3$, and $\gamma_a$, $a = 4 , \ldots , 10$, satisfy the Clifford algebras
\begin{equation}
\{\Gamma_A,\Gamma_B\}= 2\, \eta_{AB} \; , \qquad 
\{\rho_\alpha,\rho_\beta \}= 2 \, \eta_{\alpha \beta } \; , \qquad  
\{\gamma_a,\gamma_b\} = 2 \, \delta_{ab} \; ,
\end{equation}
with $\eta_{AB}$, $\eta_{\alpha \beta }$ the corresponding mostly plus Minkowski metric and $\delta_{ab}$ the Euclidean metric, and
\begin{equation}
\Gamma_0\dots\Gamma_{10}=1 \; , \qquad 
\rho_5=i\rho_0\rho_1\rho_2\rho_3 \; .
\end{equation}
Some useful relations obeyed by the $D=4$ Dirac matrices are
\begin{equation} 
	\e_{\a\b\g\d}\rho^{\d}		= -i\rho_{\a\b\g}\rho_5			\label{eq: Dirac1}	\; , \qquad 
	\e_{\a\b\g\d}\rho^{\g\d}	= -2i\rho_{\a\b}\rho_5				\; , \qquad 
	\e_{\a\b\g\d}\rho^{\b\g\d}	 =6i\rho_{\a}\rho_5		\; , 		
\end{equation}
and
\begin{equation}
	\rho_\a{}^{\d\e}=\rho^{\d\e}\rho_\a-2\rho^{[\d}\d^{\e]}_\a \; .	\label{eq: Dirac4}
\end{equation}
Finally, we use a convenient basis for the $D=11$ Dirac matrices whereby
\begin{equation}	\label{eq: CliffSplitting}
	\Gamma_\a=\rho_\a\otimes \1 \; , 
	\qquad\qquad
	\Gamma_{a}=\rho_5\otimes \gamma_a \; .
\end{equation}
%


\section{Consistency proof} \label{sec:KKproof}



\subsection{Equations of motion} \label{sec:KKproofEoms}


Assuming that the background geometry (\ref{eq: vacuum}) satisfies the $D=11$ field equations (\ref{eq: Meoms}) and imposing their $D=4$ counterparts (\ref{minimalN=2eoms}), the KK ansatz (\ref{eq:ansatz}) also solves the $D=11$ field equations provided the unknown forms $\alpha$, $\beta$ on the background geometry obey the restrictions (\ref{eq: BianchitwoformsText})--(\ref{eq:EinsteinInternalText}). Equation (\ref{eq:EinsteinExternalText}) must in turn yield the $D=4$ Einstein equation. Let us derive these equations and show that $\alpha$ and $\beta$ given in (\ref{eq:KKFormCoefs}) solve them.

In order to do this, it is convenient to split the hatted form $\hat{F}_\4$ in (\ref{Hattedg7F4}) into a background contribution, $F_\4$ in (\ref{eq: vacuum}), plus a $D=4$ graviphoton contribution using $i_\xi E_1=\norm{\xi}$:
\begin{equation}
	\hF_\4
	=F_\4-\frac g4\bar{A}\wedge i_\xi F_\4\, .
\end{equation}
The unknown forms $\alpha$ and $\beta$ can be similarly split. For calculational purposes, however, it is more convenient to sweep the $\norm{\xi}$ factors under the rug and write 
\begin{equation} \label{eq:ehatEqs}
\a=\he\wedge i_{\he^*}\alpha+\ta \; , \qquad
\b=\he\wedge i_{\he^*}\beta+\tb	 \; , \qquad \textrm{with} \quad i_{\he^*}\ta=i_{\he^*}\tb=0 \; ,
\end{equation}
where $\he \equiv \dd \psi +\mathcal{A}-g\bar{A}$ and ${\he^*}$ is the dual vector such that $i_{\he^*}\he=1$. We thus have 
\begin{equation}
\dd\a=(\dd \mathcal{A}-g \bar{F} )\wedge i_{\he^*}\alpha-\he\wedge \dd\, i_{\he^*}\alpha+\dd\ta \; ,
\end{equation}
and similarly for $\dd \beta$. With these definitions, it is now straightforward to see that $G_\4$ in (\ref{eq:ansatz}) obeys
{\setlength\arraycolsep{2pt}
\begin{eqnarray}
		\dd G_\4	
			&= &-\frac g4 \bar{F} \wedge i_\xi F_\4-\frac g4\bar{A}\wedge \dd\, i_\xi F_\4-g \bar{F} \wedge\left[(\dd \mathcal{A}-g \bar{F} )\wedge i_{\he^*}\alpha-\he\wedge \dd\, i_{\he^*}\alpha+\dd\ta\right]		\nonumber \\
			&&\quad-g\star_4 \bar{F} \wedge\left[(\dd \mathcal{A}-g \bar{F} )\wedge i_{\he^*}\beta-\he\wedge \dd\, i_{\he^*}\beta+\dd\tb\right] ,								
\end{eqnarray}
}on the $D=4$ field equations (\ref{minimalN=2eoms}) for $\bar{F}$. Imposing $\dd G_\4=0$ and requiring that the terms linear and quadratic in $\bar{F}$ and $\star_4 \bar{F}$ separately vanish, we arrive at (\ref{eq: BianchitwoformsText}). These equations imply $\alpha = \ta$, $\beta = \tb$, which we set henceforth.

We next move on to the four-form equation of motion. We fix the orientation such that $\vol_{11}=e^{11\Delta}\vol_4\wedge\vol_7$, with $\vol_4=e^0 \wedge e^1  \wedge e^2 \wedge e^3$ and \cite{Gabella:2012rc}
\begin{equation}	\label{eq: volform}
	\vol_7=-e^{4}\wedge\dots\wedge e^{10}=-E_1\wedge E_2\wedge E_3\wedge \vol (g_{\text{SU}(2)})  \;,
\end{equation}
in terms of the frame introduced in footnote \ref{eq: unwarpedelfbein}. In the following, the Hodge operators $\star_{11}$, $\star_{4}$, $\star_{7}$ are understood to be associated to the volume forms corresponding to $g_{11}$, $g_{4}$ and $g_{7}$, with $g_{4}$ the metric in \eqref{eq:ansatz} and $g_{7}$ as in the vacuum solution. With these conventions, using the torsion conditions \eqref{eq: TC1}--\eqref{eq: TC3} and the $D=4$ field equations (\ref{minimalN=2eoms}) of the graviphoton, we compute
\begin{equation} \label{eq:dStarG4}
	\begin{aligned}
		\dd\, \star_{11}G_\4	&=\vol_4\wedge \dd\, (e^{3\Delta}\star_7F_\4)-g \bar{F} \wedge\left(\frac m4\norm{\xi}e^{3\Delta} \vol (g_{\text{SU}(2)})  \wedge E_2\wedge E_3\right)	\\
					&\quad-\frac g4\star_4 \bar{F} \wedge \left[(\dd \mathcal{A}-g \bar{F} )e^{3\Delta}\wedge i_\xi\star_7\a-\he\wedge \dd\, (e^{3\Delta}\wedge i_{\xi}\star_7\a)\right]	\\
					&\quad+\frac g4 \bar{F} \wedge \left[(\dd \mathcal{A}-g \bar{F} )e^{3\Delta}\wedge i_\xi\star_7\b-\he\wedge \dd\, (e^{3\Delta}\wedge i_{\xi}\star_7\b)\right] \; .
	\end{aligned}
\end{equation}
We also find
\begin{equation} \label{eq:G4G4}
	\begin{aligned}
	G_\4\wedge G_\4	
					&=2m\vol_4\wedge F_\4-2g\hF_\4\wedge( \bar{F} \wedge\a+\star_4 \bar{F} \wedge\b)	\\
					&\quad+2g^2 \bar{F} \wedge\star_4 \bar{F} \wedge\a\wedge\b+g^2 \bar{F} \wedge  \bar{F} \wedge(\a\wedge\a-\b\wedge\b)\;.
	\end{aligned}
\end{equation}
Putting (\ref{eq:dStarG4}) and (\ref{eq:G4G4}) together, we obtain the set of equations in (\ref{eq: EomswoformsText}). 

Finally, we deal with the Einstein equation. In a frame $\{\te^A\}$ for the metric in \eqref{eq:ansatz},  $g_{11}=\eta_{AB}\te^A\otimes \te^B$, we obtain the following components of the Ricci tensor: 
\begin{align}		\label{eq:Ricci}
		\tilde{\text{Ric}}_{\a\b}	&=e^{-2\Delta}\left\{\text{Ric}_{\a\b}-\frac{g^2}{32}\norm{\xi}^2\bar{F}_{\a\g}\bar{F}{}_\b{}^\g-9(\partial_a\Delta\partial^a\Delta+\nabla_a\nabla^a\Delta)\eta_{\a\b}\right\}\;,		\nonumber \\
		\tilde{\text{Ric}}_{\a b}	&=e^{-2\Delta}\left\{-\frac g8\norm{\xi}\delta_{8b}\nabla_\g \bar{F}{}_\a{}^\g\right\}\;,				\nonumber 	\\
		\tilde{\text{Ric}}_{ab}		&=e^{-2\Delta}\Big\{\text{Ric}_{ab}+\frac{g^2}{64}\norm{\xi}^2\delta_{8a}\delta_{8b} \bar{F}_{\g\d}\bar{F}^{\g\d}	\notag \nonumber  \\
						&\quad+9\big[\partial_a\Delta\partial_b\Delta-\nabla_a\nabla_b\Delta-(\partial_c\Delta\partial^c\Delta-\nabla_c\nabla^c\Delta)\delta_{ab}\big]\Big\}\; ,
	\end{align}
where we have split the global indices $A=(\alpha,a)$ with $\alpha = 0, \ldots , 3$ and $a= 4, \ldots , 10$. In these expressions, $\text{Ric}_{\a\b}$ and $\text{Ric}_{ab}$ are the external and internal Ricci tensors in tangent space. In the same frame, the components of the four-form in \eqref{eq:ansatz} can be read off to be
\begin{equation} \label{G4comps}
	G_{\a\b\g\d}= me^{-4\Delta}\epsilon_{\a\b\g\d}\,, 				\quad
	G_{abcd}=e^{-4\Delta}F_{abcd}\,,							\quad
	G_{\a\b ab} =-ge^{-4\Delta} \big[\bar{F}_{\a\b}\a_{ab}+\tfrac12\epsilon_{\a\b\g\d}\bar{F}^{\g\d}\b_{ab} \big] ,
\end{equation}
with $\epsilon_{0123}=1$. The tangent space components, $T_{AB}\equiv \frac1{12}\left(G_{ACDE}?G_B^CDE?-\frac1{12}\eta_{AB}G^2\right)$, where $T=T_{AB}\te^A\otimes \te^B$, of the right-hand-side of the Einstein equation are thus
\begin{align}	\label{eq:SEt}
	e^{8\Delta}T_{\a\b}	&=-\frac13m^2\eta_{\a\b}+\frac{g^2}4(\a^2+\b^2)\bar{F}_{\a\g}\bar{F}_\b{}^\g-\frac{g^2}{24}\eta_{\a\b}\bar{F}^2(\a^2+2\b^2)		\nonumber \\
					&\quad-\frac{g^2}4\bar{F}_{\g(\a}?\epsilon_\b)^\g\mu\nu?\bar{F}_{\mu\nu} \, \a_{cd} \b^{cd} 
					-\frac{g^2}{24}\eta_{\a\b}\epsilon_{\mu\nu\rho\sigma}\bar{F}^{\mu\nu}\bar{F}^{\rho\sigma} \a_{cd} \b^{cd}  	\; ,	\nonumber	\\[.7em]
	e^{8\Delta}T_{\a b}	&=0	\; ,	\nonumber		\\[.7em]
	e^{8\Delta}T_{ab}	&=\frac12\left[F_{acde}?F_b^cde?-\frac1{12}\eta_{ab}F^2\right]+\frac{g^2}{24}\bar{F}^2\left[6(\a_{ac}?\a_b^c?-\b_{ac}?\b_b^c?)-\eta_{ab}(\a^2-\b^2)\right]\nonumber  \\
					&\quad+\frac{g^2}{24}\epsilon_{\mu\nu\rho\sigma}\bar{F}^{\mu\nu}\bar{F}^{\rho\sigma}\left[3(\a_{ac}?\b_b^c?+\b_{ac}?\a_b^c?)-\eta_{ab} \, \a_{cd} \b^{cd}  \right]+\frac12m^2 \, \eta_{ab} \; .
\end{align}
Equating (\ref{eq:Ricci}) and (\ref{eq:SEt}) we obtain equations (\ref{eq:EinsteinExternalText})--(\ref{eq:EinsteinInternalText}) of the main text. 

We have thus shown that the system of equations (\ref{eq: BianchitwoformsText})--(\ref{eq:EinsteinInternalText}) is equivalent to the $D=11$ Bianchi identities and equations of motion (\ref{eq: Meoms}) evaluated on the KK ansatz (\ref{eq:ansatz}), when the $D=4$ graviphoton's field equations in (\ref{minimalN=2eoms}) are imposed. Let us now verify that $\alpha$ and $\beta$ given in (\ref{eq:KKFormCoefs}) solve these equations and that, for this choice, (\ref{eq:EinsteinExternalText}) reduces to the $D=4$ Einstein equation written in (\ref{minimalN=2eoms}). The contribution in (\ref{eq: BianchitwoformsText}) that is linear in $\bar{F}$, combined with the fact that $\ta = \alpha$, implies 
$
	\dd\alpha=-\frac14i_\xi F_\4=-\frac14\dd \left(e^{3\Delta}\sqrt{1-\norm{\xi}^2}J_1\right)
$, 
where we have used (\ref{eq: undefFlux}) to compute the inner product with $\xi$. Thus, 
\begin{equation}	\label{eq: ansa}
	\a=-\frac14e^{3\Delta}\sqrt{1-\norm{\xi}^2} \, J_1+\d \; ,
\end{equation}
for a closed two-form $\delta$. As for $\beta$, we see from the torsion condition (\ref{eq: TC1}) that a natural ansatz for it that is free from legs along $E^1$ and is closed (in fact, exact), is 
\begin{equation}	\label{eq: ansb}
	\b=k \, e^{3\Delta}\left(J_3-\norm{\xi}E_2\wedge E_3\right) \; , 
\end{equation}
for some constant $k$. The forms $\alpha$, $\beta$ in (\ref{eq: ansa}), (\ref{eq: ansb}) solve, for all $\delta$ and $k$, the conditions (\ref{eq: BianchitwoformsText}) coming from the $D=11$ Bianchi identity.

The four-form equations of motion, (\ref{eq: EomswoformsText}), fix $\delta$ and $k$. First, the seven-dimensional Hodge duals of (\ref{eq: ansa}), (\ref{eq: ansb}) need to be worked out. We get
\begin{equation} \label{eq:Starsab}
	\begin{aligned}
		i_\xi\star_7\a &=\frac{\norm{\xi}}4e^{3\Delta}\sqrt{1-\norm{\xi}^2}J_1\wedge E_{2}\wedge E_{3}+i_\xi\star_7\d\;,	\\
		i_\xi\star_7\b &=-k\norm{\xi}e^{3\Delta}\left(E_{2}\wedge E_{3}-\norm{\xi}J_3\right)\wedge J_3\; .
	\end{aligned}
\end{equation}
Using (\ref{eq: ansa})--(\ref{eq:Starsab}), and (\ref{eq:dcalA}) for $d{\cal A}$, the set of equations (\ref{eq: EomswoformsText}) becomes, after some rearrangement,
\begin{align}	
	e^{6\Delta}\left\{\frac {k\norm{\xi}}4(1+4k)J_3\wedge E_2\wedge E_3+\left[\frac {\norm{\xi}^2}{32}(1+4k)+\frac12\left(k^2-\frac1{16}\right)\right]J_1\wedge J_1\right\}&	\notag\\
		+\frac12\d\wedge\left(\d-\frac12e^{3\Delta}\sqrt{1-\norm{\xi}^2}J_1\right)&=0 \; ,		\\[.7em]
	\frac{1}4\left(k+\frac14\right)e^{6\Delta}\norm{\xi}\sqrt{1-\norm{\xi}^2}J_1\wedge E_{2}\wedge E_{3}\qquad&	\notag	\\
		+e^{3\Delta}\left[\frac14i_\xi\star_7\delta+k(J_3-\norm{\xi}E_{2}\wedge E_{3})\wedge\d\right]&=0 \; , 					\\[.7em]
	me^{3\Delta}\left[-\norm{\xi}\left(\frac k2+\frac18\right)+\frac1{\norm{\xi}}\left(k+\frac14\right)\right]J_3\wedge J_3\wedge E_2\wedge E_3-\d\wedge\hF_\4\qquad&	\notag\\
		-\frac18\left(k+\frac1{4}\right)\he\wedge \dd \left[e^{6\Delta}(1-\norm{\xi}^2)J_1\wedge J_1\right]&=0 \;  ,		\\[.7em]
	\frac m{3\norm{\xi}^2}i_\xi\star_7\d\wedge\left[J_3+\left(3\norm{\xi}-\frac4{\norm{\xi}}\right)E_2\wedge E_3\right]-\frac14\he\wedge \dd \left(e^{3\Delta}i_\xi\star_7\d\right)&=0 \; .
\end{align}
It is now easy to see that all these equations are satisfied for the (very possibly, unique) choice
\begin{equation} \label{eq:deltakChoice}
\delta= 0 \; , \qquad 
k = -\tfrac14 \; .
\end{equation}

The expressions (\ref{eq:KKFormCoefs}) for $\alpha$ and $\beta$ that we brought to the main text correspond to (\ref{eq: ansa}), (\ref{eq: ansb}) with (\ref{eq:deltakChoice}). At this point we have shown that $\alpha$ and $\beta$ thus defined solve the equations (\ref{eq: BianchitwoformsText}), (\ref{eq: EomswoformsText}) implied by the Bianchi identity and equation of motion for the $D=11$ four-form. Let us see that these are also compatible with the restrictions (\ref{eq:EinsteinExternalText}), (\ref{eq:EinsteinInternalText}) implied by the $D=11$ Einstein equation. These equations can be further simplified by noting the following relation between $m$, $\Delta$ and the AdS$_4$ cosmological constant:
\begin{equation}
	9(\partial_a\Delta\partial^a\Delta+\nabla_a\nabla^a\Delta)-\tfrac13e^{-6\Delta}m^2=-12\; . 
\end{equation}
Next, reading off the tangent space components of $\alpha$, $\beta$ in (\ref{eq:KKFormCoefs}), we can compute the following contractions
\[
	\a_{ac}\b_b{}^c=-\frac1{16}\sqrt{1-\norm{\xi}^2}e^{6\Delta}\left[\delta_a^6\delta_b^5-\delta_a^7\delta_b^4+\delta_a^4\delta_b^7-\delta_a^5\delta_b^6\right]\;,
\]
\begin{equation}
	\a_{ac}\a^{bc}=\frac1{16}(1-\norm{\xi}^2)e^{6\Delta}\left[\delta_a^4\delta^b_4+\delta_a^5\delta^b_5+\delta_a^6\delta^b_6+\delta_a^7\delta^b_7\right]\;,
\end{equation}
\[
	\b_{ac}\b^{bc}=\frac1{16}e^{6\Delta}\left[\delta_a^4\delta^b_4+\delta_a^5\delta^b_5+\delta_a^6\delta^b_6+\delta_a^7\delta^b_7+\norm{\xi}^2(\delta_a^9\delta^b_9+\delta_a^{10}\delta^b_{10})\right]\; .
\]
Using these expressions, and assuming that the undeformed internal Einstein equations hold, we find that the internal components (\ref{eq:EinsteinInternalText}) of the Einstein equation vanish automatically for all values of the graviphoton $\bar{F}$. Similarly, the external components (\ref{eq:EinsteinExternalText}) of the $D=11$ Einstein equation become
\begin{equation}
	\text{Ric}_{\a\b}+12\, \eta_{\a\b}=\frac{g^2}8\left(\bar{F}_{\a\g}\bar{F}_\b{}^\g-\frac14\eta_{\a\b}\bar{F}^2\right) \; .
\end{equation} 
This coincides with the Einstein equation that derives from the $D=4$ $\cN=2$ gauged supergravity Lagrangian after a rescaling,
\begin{equation} \label{eq:4DMetResc}
\bar{g}_4=4g^{-2}\,g_4 \; , 
\end{equation}
of the four-dimensional metric.


\subsection{Supersymmetry} \label{sec:KKproofSusy}


The internal components of the $D=11$ gravitino variation (\ref{eq: susyvar}) under supersymmetry identically vanish on the KK ansatz (\ref{eq:ansatz}), and the external components reduce to the supersymmetry variations for the $D=4$ $\cN=2$ gravitino, (\ref{eq:4dGravitinoSusy}).

Let us first address the internal components. Using the gamma matrix decomposition (\ref{eq: CliffSplitting}) and the $G_\4$ components (\ref{G4comps}), some calculation allows us to write 
\begin{equation}
	\begin{aligned} \label{eq:Susy1}
		\d\Psi_a	&=\d^0\Psi_a-g\,e^{-\Delta/2}\left\{\bar{F}_{\b\g}(\rho^{\b\g}\otimes\mathbbm{1})\left[-\frac18k_a\psi_i^+\otimes\chi_i-\frac18k_a(\psi_i^+)^c\otimes\chi_i^c
				\right.\right.		\\
				&\quad+\frac{e^{-3\Delta}}{48}\a_{de}\psi_i^+\otimes?\g_a^de?\chi_i-\frac{e^{-3\Delta}}{48}\a_{de}(\psi_i^+)^c\otimes?\g_a^de?\chi_i^c				\\
				&\qquad\quad\left.-\frac{e^{-3\Delta}}{12}\a_{ae}\psi_i^+\otimes?\g^e?\chi_i+\frac{e^{-3\Delta}}{12}\a_{ae}(\psi_i^+)^c\otimes?\g^e?\chi^c_i\right]				\\
				&\quad+\bar{F}^*_{\b\g}(\rho^{\b\g}\otimes\mathbbm{1})\left[\frac{e^{-3\Delta}}{48}\b_{de}\psi_i^+\otimes?\g_a^de?\chi_i-\frac{e^{-3\Delta}}{48}\b_{de}(\psi_i^+)^c\otimes?\g_a^de?\chi_i^c\right.	\\
				&\quad\qquad\left.\left.-\frac{e^{-3\Delta}}{12}\b_{ae}\psi_i^+\otimes?\g^e?\chi_i+\frac{e^{-3\Delta}}{12}\b_{ae}(\psi_i^+)^c\otimes?\g^e?\chi^c_i\right]\right\}\;,
	\end{aligned}
\end{equation}
where we have defined $\bar{F}^*_{\d\e} \equiv \frac12\e_{\d\e\kappa\lambda}\bar{F}^{\kappa\lambda}$ and $k_a=\frac14\xi_a=\frac14\norm{\xi}\d_{a8}$. Here, $\d^0\Psi_a$ is the tensor product of $\psi^+_i$ with the left-hand-side of the first equation in (\ref{eq: KillingSpinorEq}), and thus vanishes. Using the Clifford relations (\ref{eq: Dirac1}), 
equation (\ref{eq:Susy1}) can be further simplified into
\begin{equation}
	\begin{aligned}
		\d\Psi_a&=-g\,e^{-\Delta/2}\bar{F}_{\b\g}(\rho^{\b\g}\otimes\mathbbm{1})\left[-\frac18k_a\psi_i^+\otimes\chi_i-\frac18k_a(\psi_i^+)^c\otimes\chi_i^c\right.		\\
				&+\frac{e^{-3\Delta}}{48}\a_{de}\psi_i^+\otimes?\g_a^de?\chi_i-\frac{e^{-3\Delta}}{48}\a_{de}(\psi_i^+)^c\otimes?\g_a^de?\chi_i^c	\\
				&-\frac{e^{-3\Delta}}{12}\a_{ae}\psi_i^+\otimes?\g^e?\chi_i+\frac{e^{-3\Delta}}{12}\a_{ae}(\psi_i^+)^c\otimes?\g^e?\chi^c_i			\\
				&-\frac{ie^{-3\Delta}}{48}\b_{de}\psi_i^+\otimes?\g_a^de?\chi_i-\frac{ie^{-3\Delta}}{48}\b_{de}(\psi_i^+)^c\otimes?\g_a^de?\chi_i^c	\\
				&\left.+\frac{ie^{-3\Delta}}{12}\b_{ae}\psi_i^+\otimes?\g^e?\chi_i+\frac{ie^{-3\Delta}}{12}\b_{ae}(\psi_i^+)^c\otimes?\g^e?\chi^c_i\right]\; .
	\end{aligned}
\end{equation} 
Acting with $P_\pm=\frac12(\mathbbm{1}\pm\rho_5)\otimes\mathbbm{1}$, we get that $\d\Psi_a=0$ if, and only if, the following projection holds,
\begin{equation}		\label{eq: varint}
	\Big(-6k_a+e^{-3\Delta}(\a_{de}-i\b_{de})(\g_a{}^{de}-4\d_a^d\g^e)\Big)\chi_i=0 \; ,
\end{equation}
independently for $i=1,2$. Introducing the explicit expressions (\ref{eq:KKFormCoefs}) for $\alpha$ and $\beta$, some algebra allows us to massage the relation (\ref{eq: varint}), for $a=8$, into (\ref{eq: cond8}) and, for $a \neq 8$, into (\ref{eq: proj3}) of the main text. These projections can be checked to be fully compatible with the SU(2)--structure that defines the background geometry, without giving independent restrictions on the Killing spinors $\chi_i$. As an instance of how this works, the projector \eqref{eq: cond8} gives rise to a bilinear
\begin{align}
	\bar{\chi}_+^c&\left[\norm{\xi}(3\g^8+i\g^{910})+\sqrt{1-\norm{\xi}^2}(\g^{46}-\g^{57})-i(\g^{45}+\g^{67})\right]\chi_-	\notag	\\
		&=\norm{\xi}\Big(3(-i\norm{\xi})+i\norm{\xi}\Big)+\sqrt{1-\norm{\xi}^2}\big(-2i\sqrt{1-\norm{\xi}^2}\big)-i(-2)\;,
\end{align}
with $\chi_\pm=\frac1{\sqrt2}(\chi_1\pm i\chi_2)$, and where we have used (B.2), (B.3) of \cite{Gabella:2012rc}. This vanishes identically.

Next, we turn to the external variations of the gravitino. Particularising (\ref{eq: susyvar}) to external indices, employing the basis (\ref{eq: CliffSplitting}) for the Dirac matrices, and extensively using the underformed Killing spinor equations (\ref{eq: KillingSpinorEq}), we can write
\begin{equation} \label{eq:SusyGravitinoExt1}
	\begin{aligned}
		\d\Psi_\mu
				& = 
					e^{\Delta/2}\left\{\grad_\mu\psi^+_i\otimes\chi_i-\rho_\mu\psi^+_i\otimes\chi^c_i-\frac{g\norm{\xi}}{16}\bar{F}_{\mu \b}\rho^\b\psi^+_i\otimes\gamma^8\chi_i\right.	
					\\
				&\quad+\frac g4\grad_bk_c\bar{A}_\mu\psi^+_i\otimes\gamma^{bc}\chi_i+\frac{g\norm{\xi}}4 \bar{A}_{\mu}\psi^+_i\otimes\grad_8\chi_i
					\\
				&\quad-\frac{g^2\norm{\xi}^2}{128} \bar{A}_{\mu}\bar{F}_{\b\g}\rho^{\b\g}\psi^+_i\otimes\chi_i
					-\frac{ge^{-3\Delta}}{48}\left(\bar{F}_{\d\e}\a_{bc}+\bar{F}^*_{\d\e}\b_{bc}\right)?\rho_\mu^\d\e?\psi^+_i\otimes\gamma^{bc}\chi_i					\\
				&\quad+\frac{g^2\norm{\xi}e^{-3\Delta}}{192}\bar{A}_\mu\left(\bar{F}_{\d\e}\a_{bc}+\bar{F}^*_{\d\e}\b_{bc}\right)?\rho^{\d\e}?\psi^+_i\otimes?\g_8^bc?\chi_i	\\
				&\quad+\left.\frac{ge^{-3\Delta}}{12}\left(\bar{F}_{\mu\g}\a_{de}+\bar{F}^*_{\mu\g}\b_{de}\right)\rho^\g\psi^+_i\otimes\g^{de}\chi_i\right\}+m.c.
	\end{aligned}
\end{equation} 
From $(2.24)$ of \cite{Gabella:2012rc} and $\mathcal{L}_\xi\chi=\grad_\xi\chi+\frac14\grad_a\xi_b\gamma^{ab}\chi$ (see \cite{Ortin:2002qb}),  we find that $\mathcal{L}_\xi\chi_{1}=-2\chi_2$ and $\mathcal{L}_\xi\chi_{2}=2\chi_1$, so that 
\begin{equation}
\norm{\xi}\grad_8\chi_1+\grad_ak_b\gamma^{ab}\chi_1=-2\chi_2 \; , \qquad 
\norm{\xi}\grad_8\chi_2+\grad_ak_b\gamma^{ab}\chi_2=2\chi_1 \; .
\end{equation}
Bringing these relations to (\ref{eq:SusyGravitinoExt1}) and using the $D=4$ Dirac matrix relations (\ref{eq: Dirac1}), (\ref{eq: Dirac4}) to get rid of the $\bar{F}^*_{\d\e}$ terms, we obtain
\begin{align} \label{eq:SusyGravitinoExt2}
	\d\Psi_\mu
			&=e^{\Delta/2}\left\{\grad_\mu\psi^+_i\otimes\chi_i-\rho_\mu(\psi^+_i)^c\otimes\chi_i-\frac{g\norm{\xi}}{16}\bar{F}_{\mu \b}\rho^\b\psi^+_i\otimes\gamma^8\chi_i	
			-\frac {ig}2\e_{ij}\bar{A}_\mu\psi^+_i\otimes\chi_j\right.	\notag\\
			&\left.\quad-\frac{ge^{-3\Delta}}{48}\bar{F}_{\d\e}\Big[\a_{bc}\big(\rho^{\d\e}\rho_\mu+2\rho^{\d} e^{\e}{}_\mu\big)\psi^+_i
				+2i\b_{bc}\big(\rho^{\d\e}\rho_\mu-\rho^{\d} e^{\e}{}_\mu\big)\psi^+_i	\Big]\otimes\g^{bc}\chi_i\right\}+m.c. \;,
\end{align}
where $e^{\e}{}_\mu$ are the frame components. We can now use the $G$-structure compatible projections (\ref{eq: cond8}), (\ref{eq: proj3}) to further simplify the result. Using them, (\ref{eq:SusyGravitinoExt2}) becomes
\begin{align}  \label{eq:SusyGravitinoExt3}
	\d\Psi_\mu
			=e^{\Delta/2}&\left\{\grad_\mu\psi^+_i\otimes\chi_i-\rho_\mu(\psi^+_i)^c\otimes\chi_i-\frac {ig}2\e_{ij}\bar{A}_\mu\psi^+_i\otimes\chi_j\right.		\notag\\
			&\left.+\frac{ig}{16}\bar{F}_{\d\e}\,\rho^{\d\e}\rho_\mu\,\psi_i^+\otimes\g^{45}\chi_i\right\}+m.c.
\end{align}
At this point, we recognise one more projection, (\ref{eq:PrjGravExt}) of the main text, that may be imposed to relate the internal spinors $\chi_i$ to their charge conjugates $\chi_i^c$. This projection is, again, fully compatible with the original Killing spinor equations (\ref{eq: KillingSpinorEq}) and does not constrain the background geometry any further. Using (\ref{eq:PrjGravExt}) along with $(\chi_i^c)^c=\chi_i$ and $(\rho_{\sst{(n)}}\psi^+_i)^c=\rho_{\sst{(n)}}(\psi^+_i)^c$, equation (\ref{eq:SusyGravitinoExt3}) finally yields 
\begin{equation} \label{eq:SusyGravitinoExt4}
	\d\Psi_\mu =e^{\Delta/2}\left\{\grad_\mu\psi^+_i-\rho_\mu(\psi^+_i)^c+\frac {ig}2\e_{ij}\bar{A}_\mu\psi^+_j +\frac{g}{16}\bar{F}_{\d\e}\,\rho^{\d\e}\rho_\mu\,\e_{ij}(\psi_j^+)^c\right\}\otimes\chi_i+m.c.
\end{equation}
If the external components $\Psi_\mu$ of the $D=11$ gravitino and the $D=4$ gravitini $\psi _{i\mu}^+$ are related as in equation (\ref{eq:KKGravitino}) of the main text, then (\ref{eq:SusyGravitinoExt4}) reproduces the supersymmetry variations (\ref{eq:4dGravitinoSusy}) for the gravitini of $D=4$ $\cN=2$ supergravity, after the metric rescaling (\ref{eq:4DMetResc}) is taken into account.

\bibliography{references}

\providecommand{\href}[2]{#2}\begingroup\raggedright\begin{thebibliography}{10}

\bibitem{Gauntlett:2007ma}
J.~P. Gauntlett and O.~Varela, {\it {Consistent Kaluza-Klein reductions for
  general supersymmetric AdS solutions}},  {\em Phys.Rev.} {\bf D76} (2007)
  126007, [\href{http://arxiv.org/abs/0707.2315}{{\tt arXiv:0707.2315}}].

\bibitem{Duff:1985jd}
M.~J. Duff and C.~N. Pope, {\it {Consistent truncations in Kaluza-Klein
  theories}},  {\em Nucl. Phys.} {\bf B255} (1985) 355--364.

\bibitem{Pope:1987ad}
C.~N. Pope and K.~S. Stelle, {\it {Zilch Currents, Supersymmetry and
  {Kaluza-Klein} Consistency}},  {\em Phys. Lett.} {\bf B198} (1987) 151.

\bibitem{Gauntlett:2002sc}
J.~P. Gauntlett, D.~Martelli, S.~Pakis, and D.~Waldram, {\it {G structures and
  wrapped NS5-branes}},  {\em Commun. Math. Phys.} {\bf 247} (2004) 421--445,
  [\href{http://arxiv.org/abs/hep-th/0205050}{{\tt hep-th/0205050}}].

\bibitem{Cremmer:1978km}
E.~Cremmer, B.~Julia, and J.~Scherk, {\it {Supergravity Theory in
  Eleven-Dimensions}},  {\em Phys.Lett.} {\bf B76} (1978) 409--412.

\bibitem{Fradkin:1976xz}
E.~S. Fradkin and M.~A. Vasiliev, {\it {Model of Supergravity with Minimal
  Electromagnetic Interaction}}, .

\bibitem{Freedman:1976aw}
D.~Z. Freedman and A.~K. Das, {\it {Gauge Internal Symmetry in Extended
  Supergravity}},  {\em Nucl. Phys.} {\bf B120} (1977) 221--230.

\bibitem{Pope:1985jg}
C.~N. Pope, {\it {Consistency of truncations in Kaluza-Klein}},  {\em Conf.
  Proc.} {\bf C841031} (1984) 429--431.

\bibitem{Freund:1980xh}
P.~G. Freund and M.~A. Rubin, {\it {Dynamics of Dimensional Reduction}},  {\em
  Phys.Lett.} {\bf B97} (1980) 233--235.

\bibitem{Duff:1984hn}
M.~J. Duff, B.~E.~W. Nilsson, C.~N. Pope, and N.~P. Warner, {\it {On the
  Consistency of the {Kaluza-Klein} Ansatz}},  {\em Phys. Lett.} {\bf 149B}
  (1984) 90--94.

\bibitem{Gauntlett:2006ux}
J.~P. Gauntlett, O.~A.~P. Mac~Conamhna, T.~Mateos, and D.~Waldram, {\it {AdS
  spacetimes from wrapped M5 branes}},  {\em JHEP} {\bf 11} (2006) 053,
  [\href{http://arxiv.org/abs/hep-th/0605146}{{\tt hep-th/0605146}}].

\bibitem{Gabella:2012rc}
M.~Gabella, D.~Martelli, A.~Passias, and J.~Sparks, {\it {${\cal N}=2$
  supersymmetric AdS$_{4}$ solutions of M-theory}},  {\em Commun. Math. Phys.}
  {\bf 325} (2014) 487--525, [\href{http://arxiv.org/abs/1207.3082}{{\tt
  arXiv:1207.3082}}].

\bibitem{Tsikas:1986rx}
T.~T. Tsikas, {\it {Consistent Truncations of Chiral $N=2 D=10$ Supergravity on
  the Round Five Sphere}},  {\em Class. Quant. Grav.} {\bf 3} (1986) 733.

\bibitem{Buchel:2006gb}
A.~Buchel and J.~T. Liu, {\it {Gauged supergravity from type IIB string theory
  on Y**p,q manifolds}},  {\em Nucl.Phys.} {\bf B771} (2007) 93--112,
  [\href{http://arxiv.org/abs/hep-th/0608002}{{\tt hep-th/0608002}}].

\bibitem{Gauntlett:2006ai}
J.~P. Gauntlett, E.~O~Colgain, and O.~Varela, {\it {Properties of some
  conformal field theories with M-theory duals}},  {\em JHEP} {\bf 02} (2007)
  049, [\href{http://arxiv.org/abs/hep-th/0611219}{{\tt hep-th/0611219}}].

\bibitem{Gauntlett:2007sm}
J.~P. Gauntlett and O.~Varela, {\it {D=5 SU(2) x U(1) Gauged Supergravity from
  D=11 Supergravity}},  {\em JHEP} {\bf 02} (2008) 083,
  [\href{http://arxiv.org/abs/0712.3560}{{\tt arXiv:0712.3560}}].

\bibitem{Passias:2015gya}
A.~Passias, A.~Rota, and A.~Tomasiello, {\it {Universal consistent truncation
  for 6d/7d gauge/gravity duals}},  {\em JHEP} {\bf 10} (2015) 187,
  [\href{http://arxiv.org/abs/1506.05462}{{\tt arXiv:1506.05462}}].

\bibitem{Malek:2017njj}
E.~Malek, {\it {Half-Maximal Supersymmetry from Exceptional Field Theory}},
  {\em Fortsch. Phys.} {\bf 65} (2017), no.~10-11 1700061,
  [\href{http://arxiv.org/abs/1707.00714}{{\tt arXiv:1707.00714}}].

\bibitem{Hong:2018amk}
J.~Hong, J.~T. Liu, and D.~R. Mayerson, {\it {Gauged Six-Dimensional
  Supergravity from Warped IIB Reductions}},  {\em JHEP} {\bf 09} (2018) 140,
  [\href{http://arxiv.org/abs/1808.04301}{{\tt arXiv:1808.04301}}].

\bibitem{Malek:2018zcz}
E.~Malek, H.~Samtleben, and V.~Vall~Camell, {\it {Supersymmetric AdS$_{7}$ and
  AdS$_6$ vacua and their minimal consistent truncations from exceptional field
  theory}},  {\em Phys. Lett.} {\bf B786} (2018) 171--179,
  [\href{http://arxiv.org/abs/1808.05597}{{\tt arXiv:1808.05597}}].

\bibitem{Liu:2019cea}
J.~T. Liu and B.~McPeak, {\it {Gauged Supergravity from the Lunin-Maldacena
  background}},  \href{http://arxiv.org/abs/1905.06861}{{\tt
  arXiv:1905.06861}}.

\bibitem{Cassani:2019vcl}
D.~Cassani, G.~Josse, M.~Petrini, and D.~Waldram, {\it {Systematics of
  consistent truncations from generalised geometry}},
  \href{http://arxiv.org/abs/1907.06730}{{\tt arXiv:1907.06730}}.

\bibitem{Gauntlett:2002fz}
J.~P. Gauntlett and S.~Pakis, {\it {The Geometry of D = 11 killing spinors}},
  {\em JHEP} {\bf 04} (2003) 039,
  [\href{http://arxiv.org/abs/hep-th/0212008}{{\tt hep-th/0212008}}].

\bibitem{Corrado:2001nv}
R.~Corrado, K.~Pilch, and N.~P. Warner, {\it {An N=2 supersymmetric membrane
  flow}},  {\em Nucl. Phys.} {\bf B629} (2002) 74--96,
  [\href{http://arxiv.org/abs/hep-th/0107220}{{\tt hep-th/0107220}}].

\bibitem{Larios:2019kbw}
G.~Larios, P.~Ntokos, and O.~Varela, {\it {Embedding the SU(3) sector of SO(8)
  supergravity in $D=11$}},  \href{http://arxiv.org/abs/1907.02087}{{\tt
  arXiv:1907.02087}}.

\bibitem{Varela:2015ywx}
O.~Varela, {\it {Complete $D=11$ embedding of SO(8) supergravity}},  {\em Phys.
  Rev.} {\bf D97} (2018), no.~4 045010,
  [\href{http://arxiv.org/abs/1512.04943}{{\tt arXiv:1512.04943}}].

\bibitem{Halmagyi:2012ic}
N.~Halmagyi, K.~Pilch, and N.~P. Warner, {\it {On Supersymmetric Flux Solutions
  of M-theory}},  \href{http://arxiv.org/abs/1207.4325}{{\tt arXiv:1207.4325}}.

\bibitem{Aharony:2008ug}
O.~Aharony, O.~Bergman, D.~L. Jafferis, and J.~Maldacena, {\it {N=6
  superconformal Chern-Simons-matter theories, M2-branes and their gravity
  duals}},  {\em JHEP} {\bf 10} (2008) 091,
  [\href{http://arxiv.org/abs/0806.1218}{{\tt arXiv:0806.1218}}].

\bibitem{Caldarelli:2003pb}
M.~M. Caldarelli and D.~Klemm, {\it {All supersymmetric solutions of N=2, D = 4
  gauged supergravity}},  {\em JHEP} {\bf 09} (2003) 019,
  [\href{http://arxiv.org/abs/hep-th/0307022}{{\tt hep-th/0307022}}].

\bibitem{Cacciatori:2004rt}
S.~L. Cacciatori, M.~M. Caldarelli, D.~Klemm, and D.~S. Mansi, {\it {More on
  BPS solutions of N = 2, D = 4 gauged supergravity}},  {\em JHEP} {\bf 07}
  (2004) 061, [\href{http://arxiv.org/abs/hep-th/0406238}{{\tt
  hep-th/0406238}}].

\bibitem{Ortin:2002qb}
T.~Ortin, {\it {A Note on Lie-Lorentz derivatives}},  {\em Class. Quant. Grav.}
  {\bf 19} (2002) L143--L150, [\href{http://arxiv.org/abs/hep-th/0206159}{{\tt
  hep-th/0206159}}].

\end{thebibliography}\endgroup

\end{document}